\documentclass{ws-procs975x65}
\usepackage{graphicx}
\usepackage{amsmath}
\usepackage{amssymb}
\usepackage{enumerate}
\usepackage{epsfig}
\usepackage{verbatim}
\usepackage{array}

\makeatletter

\def\vereq#1#2{\lower3pt\vbox{\baselineskip0.5pt\lineskip0.5pt
\ialign{$\m@th#1\hfill##\hfil$\crcr#2\crcr\sim\crcr}}}

\makeatother

\begin{document}

\title{The Pauli Exclusion Principle, Spin, and Statistics in Loop Quantum
       Gravity: $SU(2)$ versus $SO(3)$\footnote{
  \uppercase{T}alk at
  the 10th \uppercase{M}arcel \uppercase{G}rossmann \uppercase{M}eeting
  in \uppercase{R}io de \uppercase{J}aneiro,
  20--26 \uppercase{J}uly 2003}}

\author{John Swain}
\address{
Department of Physics,
Northeastern University, Boston, MA 02115\\
e-mail: {\tt john.swain@cern.ch}}


\maketitle

\abstracts{Recent attempts to resolve the ambiguity in the loop quantum
gravity description of the quantization of area has led to 
the idea that $j=1$ edges of spin-networks dominate in 
their contribution to black hole areas as opposed to $j=1/2$ which
would naively be expected. This suggests that
the true gauge group involved might be $SO(3)$ rather than $SU(2)$.
We argue that the idea that a
version of the Pauli principle is present in loop quantum gravity
allows one to maintain $SU(2)$ as the gauge group while still 
naturally achieving the desired suppression of spin-1/2 punctures.
Such an idea can be motivated by arguments from geometric quantization
even though the $SU(2)$ under consideration does not have the geometrical
interpretation of rotations in 3-dimensional space, and its representation
labels do not correspond to physical angular momenta. In this picture,
it is natural that macroscopic areas come almost entirely 
from $j=1$ punctures rather than $j=1/2$ punctures, and this is
for much the same reason that photons lead to macroscopic classically
observable fields while electrons do not.}

\section{Introduction}
This talk is based on an essay which received an 
an honorable mention in the 2003 Essay Competition of the 
Gravity Research Foundation\cite{grgpaper}, and its subsequent
and significant amplification\cite{IJMPDpaper}.

It is a contribution to trying to figure out what the rules should
be for a theory of quantum gravity in 3+1 dimensions based on
a nonperturbative treatment in the framework now generally
referred to as ``loop quantum gravity''.
The successes of this approach
using the Ashtekar variables have been numerous and significant
and include proofs that area and volume operators have discrete
spectra, and a derivation of black hole entropy up to an overall, yet
undetermined constant\cite{Ashtekarstuff}. An excellent recent review
leading directly to this paper is by Baez\cite{Baez}, and
its influence on this introduction will be clear.

One thinks of a spin-network as floating freely, but able to intersect
physical surfaces, with each edge contributing to its area.
To a good approximation, the area $A$ of a surface which intersects
a spin network at $i$ edges, each carrying an $SU(2)$ label $j$ is given
in geometrized units (Planck length equal to unity) by 
$A \approx \sum_i 8\pi\gamma\sqrt{j_i(j_i+1)}$,
where $\gamma$ is the Immirzi-Barbero parameter\cite{Immirzi-Barbero}.
The most important microstates 
consistent with a given area are those for which $j$ is as small as possible,
which one would expect to be $j_{min}=1/2$. In this case, each contribution
to the area corresponds to a spin $j=1/2$ which can come in two possible
$m$ values of $\pm 1/2$. For $n$ punctures, we have
$A \approx 4\pi\sqrt{3}\gamma n$ and entropy $S\approx \ln(2^n) \approx
\frac{\ln(2)}{4\pi\sqrt{3}\gamma}A$. 

Now if we use
Hawking's formula\cite{Hawking} for black hole entropy $S=A/4$ to get
$\gamma=\frac{\ln(2)}{\pi\sqrt{3}}$ and the smallest quantum of 
area is then $8\pi\gamma\sqrt{\frac{1}{2}(\frac{1}{2}+1)} = 4\ln(2)$.
A black hole's horizon then acquires area, to a good approximation, 
from the punctures of many
spin network edges, each carrying a quantum of area $4\ln(2)$ and
one ``bit'' of information, in accordance with Wheeler's ``it from bit''
philosophy\cite{Wheeler}.

Bekenstein's early intuition\cite{Beckenstein} that the area operator
for black holes should have a discrete spectrum made of equal area steps
(something not really quite true in loop quantum gravity in full generality)
was followed by Mukhanov's observation \cite{Mukhanov} that 
the $n^{th}$ area state should have degeneracy $k^n$ with
steps between areas of $4\ln(k)$ for $k$ some integer $\geq 2$ 
in order to reproduce Hawking's expression $S=A/4$. For $k=2$ one would have
the $n^{th}$ area state described by $n$ binary {\em bits}. 

On the other hand Hod\cite{Hod} has argued that by looking at the
quasinormal damped modes\cite{Cardoso} of a classical back hole one should be
able to derive the quanta of area using 
the formula $A=16\pi M^2$ relating area and mass of a
black hole to get $\Delta A=32\pi M \Delta M$ for the change in area 
accompanying an emission of energy $\Delta M$. Nollert's computer
calculations\cite{Nollert}
of the asymptotic frequency $\omega$ of the damped normal modes gave
$\omega \approx 0.4371235/M$, so setting $\omega = \Delta M$ one finds
$\Delta A \approx 4.39444$, which one might guess is exactly $4\ln(3)$. 
Motl \cite{Motl} showed that this is indeed the case.

One might expect\cite{Dreyer}
$\Delta A \approx 4\ln(3)$ instead of $\Delta A \approx 4\ln(2)$
if the spin network edges contributing to the area of a black
hole didn't carry $j=1/2$, but rather $j=1$. In this case
$j_{min}$ would be $1$ rather than $1/2$,
there would be
three possible $m$ values, area elements would be described not by
binary ``bits'', but by trinary ``trits''(see also \cite{MotlandNeitzke})
and it suggests
that the correct gauge group might be $SO(3)$ rather than $SU(2)$.

So which is it? Corichi has argued\cite{Corichi} that one might arrive at
the conclusion that $j_{min}=1$ by thinking of
a conserved fermion number being assigned to each spin-1/2 edge.
The exact physical interpretation of this is not entirely clear, but
roughly one might imagine that there is a sort of conserved fermion
number which would disallow a single spin network edge from leaving
a surface and being left ``dangling in the bulk''. Ling and Zhang\cite{LiZhang} have
argued that going to $N=1$ supergravity would also offer a way to
avoid spin-1/2 edges.

\section{An Exclusion Principle?}

The point of the original essay for the Gravity Research Foundation was
to  suggest that if one assumes (following Pauli and real spins)
that no more than one $j=1/2$ edge (think of ``one spin-1/2 particle'')
would allowed to puncture a surface (think of ``occupy a given quantum state''),
then everything makes sense. How can one justify such a conjecture?

In quantum mechanics, the spin-statistics theorem is simply a postulate:
as Dirac\cite{Dirac} puts it ``to get agreement with experiment one
must {\em assume} that two electrons are never in the same state'' 
(my italics). If one is honest, the usual quantum field theory arguments
are not much better.
The requirement that field operators should commute at spacelike separations 
which seems reasonable if one wants appropriate notions of causality and
locality comes to grief for fermions ({\em i.e.} one has an unstable
vacuum). Jordan's anticommutators \cite{Jordan}
save the day, but obscure the connection to causality unless one
assumes that the fermion fields are Grassmann-valued. 
A rather comprehensive
review of the history and literature is in \cite{Paulibook}.

For a surface punctured by spin network edges I want to argue that one should
consider an amplitude which returns to its original value, up to 
a phase, upon the exchange of two spin-1/2 (and thus identical,
indistinguishable) punctures. If making the exchange twice leads to 
the identity, one then needs merely to
choose a sign, and $-1$ seems at least as natural as $+1$. This argument
can be sharpened in the following way:

For a configuration space of $n$ spin-$j$ non-coincident
{\em identical} punctures,
as shown long ago by Laidlaw and DeWitt\cite{LaidlawDeWitt},
phases for propagators 
must form a scalar unitary representation of the fundamental group.
This limits the possible choice of statistics 
to Bose statistics (no phase change under permutations) 
or Fermi statistics (change of sign for any odd permutation).
As it stands, this is just
a statement about what statistics are possible, but
has nothing to do with rotations, $SU(2)$, $SO(3)$, or even physical space.

Now to argue for Fermi statistics for odd half-integer $j$ punctures
and Bose statistics for integer $j$ punctures we cannot use the usual
QFT arguments as we have no background spacetime. The same applies to
heuristic arguments based on identifying an exchange as a
composition of physical rotations ({\it i.e.} \cite{Donth}) and to
arguments based on extended kink-like objects ({\it i.e.} \cite{kinks}),
since the punctures are meant to be points. Related approaches
such as those of Balachandran {\it et al.} \cite{Balachandran},
Tseuschner\cite{Tscheuschner}, Berry and Robbins\cite{Berry}, and many others
in \cite{Paulibook} all seem difficult to apply.
For an argument not dependent on
a prior concept of physical space
we look directly at the configuration space and use
ideas from geometric quantization\cite{Geometricq},
following beautiful arguments
of Anastopoulos\cite{Anastopoulos}.


Consider the quantization of the sphere 
$S^2=\lbrace (x_1,x_2,x_3) | x_1^2+x_2^2+x_3^2=1 \rbrace$
with the symplectic form
$\Omega = \frac{1}{2} s \epsilon_{ijk} x^i dx^j\wedge dx^k$
and a symplectic action of $SO(3)$ on $S^2$ where $SO(3)$ acts on 
the $x^i$ in the usual way by its defining representation and
obviously leaves $\Omega$ invariant.
Each choice of $s$ gives a different symplectic manifold,
and the requirement that $\Omega$ be
integrable requires that $s=n/2$ with $n$ an integer.
In this way $s$ corresponds to
the usual notion of spin in quantum mechanics.
Note that so far there is no explicit identification of the $x^i$ with
spacetime directions -- they just happen to define the coordinates on
an {\em abstract} $S^2$. We also have 
the $U(1)$ bundle provided by the Hopf map from $S^3$ to $S^2$ defined
by $\pi(\xi)^i = \bar{\xi}\sigma^i\xi$,
defined in terms of 2-component complex ``spinors'' $\xi$ of length 1 and
$\sigma^i$ the Pauli matrices. $\pi(\xi)^i$ is obviously real
and invariant under $\xi\rightarrow u\xi$ for any $u=\exp(i\theta)$ in
$U(1)$.  For $s=1/2$  we have $U(1)$ operations corresponding
to the two square roots of unity $\pm 1$, {\em both} of which correspond
to the identity element of $SO(3)$.

A prequantization is given by 
the bundle whose total space is the set
of orbits of $S^3$ under the two $U(1)$ actions of multiplication
by $\pm 1$ and the same projection map $\pi$. 
$SU(2)$ actions on $S^3$ correspond to $SO(3)$ actions on $S^2$, and we pick
up a factor of $-1$ for a $2\pi$ rotation. We can now think
of an $s=1/2$ state as a point on the sphere $S^2$, accompanied
by this sign change for $2\pi$ rotations.
Now if we make this construction twice out of two classical phase
spaces, but assume that the two are {\em the same} $S^2$, we can
exchange two points on the $S^2$ (``two spins'') by two $SO(3)$ rotations,
one acting on each point in {\em the same} $S^2$.  For spin-1/2 this gives
a factor of $-1$ on exchange, and for general spin-$j$ the expected
spin-statistics connection\cite{IJMPDpaper}.

In other words, even though one might colloquially speak
of the edges as carrying ``spins'', knowing full well that this is really
a way of saying ``$SU(2)$ representation labels with no obviously necessary
connection to spin of elementary particles or irreducible representations
of the rotation group in physical space'', in fact it {\em does} make
sense to think of them as physical spins and argue for a
spin-statistics theorem. In this sense the spin-statistics theorem might
better be thought of as a sort of
``($SU(2)$ representation label)-(sign change or not on exchange)''theorem.

In loop quantum gravity this leads
then to a picture in which a large black hole {\em does} receives 
contributions to its area from
spin-1/2 (and spin-3/2, spin-5/2, {\em etc.}) punctures, but these
are always very small compared to the enormous number of $j=1$
edges. The value $j=1$ is the lowest value of $j$ contributing
nonzero area not
being severely limited by Fermi-Dirac statistics, and able to 
appear arbitrarily often. This can then make it look like we're dealing
with $SO(3)$ rather than $SU(2)$.

\section{Conclusions}

In a sense, the question of $SU(2)$ vs. $SO(3)$ in 
loop quantum gravity could be
very much like one that we face in everyday physics. Integer spin particles,
which fall into $SO(3)$ representations, obey Bose-Einstein statistics and
gregariously bunch together to give large macroscopically observable
fields such as electromagnetic fields.  Half-integer spin particles do not.
We could well be excused
for thinking 
that the symmetry group of our world under rotations was $SO(3)$
rather than $SU(2)$. Indeed, until the discovery of spin, it did appear
that physical rotations were always elements of $SO(3)$. The
need for $SU(2)$ was, in many ways, a surprise!

These ideas are likely hard to test, 
but it is possible to make some predictions.
For example, the $SU(2)$ theory with the exclusion principle proposed
here will give both: a) what seems to be the correct result for large
black holes, with areas well-described by values which go up {\em in
steps of} $4\ln(3)$; {\em and}
b) the possibility of
simultaneously allowing areas {\em as small as} $4\ln(2)$.

\vspace*{-3mm}
\section*{Acknowledgments}

It is a great pleasure to thank the session and  conference organizers, especially
Carlo Rovelli (via email),
Robert Oeckl, Santiago Perez Bergliaffa, Mario Novello, and Remo Ruffini
for their help and hospitality, and
everyone in Rio, the most beautiful city in the world! Muito obrigado!
\vspace*{-3mm}
\small

\end{document}